\documentstyle[preprint,prl,aps]{revtex}

\begin{document}

\draft

\title{Experimental Verifications of the Casimir Attractive Force
between Solid Bodies}

\author{S.K. Lamoreaux}

\address{University of California,
Los Alamos National Laboratory,
Physics Division 23, M.S. H803, Los Alamos, NM 87545}

\date{\today}

\maketitle

\noindent
(Submitted for the Casimir 90th Birthday Issue of Comments on
Atomic and Molecular Physics)
\tightenlines

\begin{abstract}

A brief review of the recent experimental verifications of the Casimir force 
between extended bodies is presented.  With modern techniques, it
now appears feasible to test the force law with precision better
than 1\%;  I will address the issues relating to the theoretical
interpretation of experiments at this level of accuracy.

\bigskip
\noindent
Keywords:  Casimir Force, Quantum Electrodynamics, Fluctuations\hfill\break


\end{abstract}

\section{Introduction}

The force between uncharged conducting surfaces, the so-called ``Casimir
force,'' has been described as one of the least intuitive consequences
of quantum electrodynamics.
For conducting parallel flat plates separated
by distance $r$, this force per unit area $A$
has the magnitude \cite{cas}:
\begin{equation}\label{cf}
F(d)/A={\pi^2\over 240}{\hbar c\over d^4}=0.013 {1\over d^4}
{\rm dyn (\mu m)^4/cm^2}.
\end{equation}
This relationship can be derived
by considering the electromagnetic mode structure
between the two plates, as compared to free space, and by assigning 
a zero-point energy of ${1\over 2}\hbar\omega$ to each 
electromagnetic mode (photon).  The change
in total energy density between the plates, as compared to free
space, as a function of separation $d$, leads to the force of attraction.

The only fundamental constants that enter Eq. (1) are $\hbar$ and
$c$; the electron charge $e$ is absent, implying that the 
electromagnetic field is not coupling to matter.  The role of
$c$ is to convert the electromagnetic
mode wavelength to a frequency, while
$\hbar$ converts the frequency to an energy.

The term ``Casimir effect'' is applied to a number of
long-range interactions, such as those between atoms or molecules
(retarded van der Waals interaction), an atom and a material
surface (Casimir-Polder interaction), and the attraction between
bulk material bodies. The latter effect is generally referred to 
as ``the'' Casimir force and depends only on the bulk properties of
the bodies under consideration; I will limit my discussion
to the latter effect.

For real materials, Eq. (1) must break down when the separation
$d$ is so small that the mode frequencies are higher than the
plasma frequency (for a metal) or higher than the absorption
resonances (for a dielectric) of the material used to make the plates;
for sufficiently small separation, the
force of attraction varies as $1/d^3$, as discussed
in particular by Lifshitz \cite{lif}.  In analogy with the attractive
forces between atoms, the force in this range is sometimes
referred to as the London-van der Waals attraction, while
the $1/d^4$ range is referred to as the retarded van der Waals 
(Casimir) interaction.  For the Casimir effect,  
the crossover distance between the
regimes is  $d\sim100$ nm, much larger than atomic spacings in
the materials,  so it still makes sense to describe the materials
by their bulk properties (index of refraction); the $1/d^3$
vs. $1/d^4$ interaction is in this case due to the truncation of the mode
frequencies that are affected by the changing plate separation.  
Therefore the  crossover between the two regimes appears to be
of physically different origin
than in the case of the attractive forces between isolated
atoms.  

The Casimir effect and its calculation represent an
electromagnetic waveguide problem, where imperfect materials
are used in the construction of the waveguide.  The zero-point
fields are those associated with the waveguide; these modes do
not exist in free space, so the idea that the Casimir effect
represents a negative energy density compared to free space 
is incorrect.

Given that the distances where the force of
attraction is sufficiently strong to be experimentally
detected are $d\sim 1000$ nm or less,
an accurate theoretical description of an experimental system
must take into account the real material
properties as will be discussed in Sec. III below.

Finally, it is interesting to note that there have been only a dozen
or so published experimental measurements of the Casimir force, to be compared
with the hundreds and hundreds of theoretical papers on the subject.
Perhaps very few doubt the strict validity of Eq. (1) or its
modification for real materials \cite{lif}.  Because of the unavoidable 
uncertainties in bulk material and surface properties, verification 
of Eq. (1) as a test of QED will likely always be inferior to 
measurements of the Lamb shift or $g-2$ of the electron.

\section{Experiments}

\subsection{Overview}

The experimental situation as of 1989 has been reviewed by Sparnaay
in  the volume prepared in the honor of Dr. Casimir's 80th 
birthday \cite{spar}.  Since then, two experiments have been
performed, both with significantly better accuracy than had
been previously obtained.  These two experiments were based
on a torsion pendulum balance \cite{lam} and  on atomic force
microscopy (AFM) \cite{moh}.  

The recent experiments employed techniques that were  developed in particular
by van Blokland and Overbeek \cite{over} in the measurement of the attractive
forces between metallic films.  Measurements between metallic films
pose difficult problems as compared to dielectric films for which
optical techniques can be used for alignment and distance measurements.
In the case of metallic films, the distance is determined by
measurement of the capacitance between the plates.  Alignment is
simplified by making one plate convex, in which case the
geometry is fully determined by the radius of curvature at
the point of closest approach, and the distance between the
plates at that point.  This technique was first put forward by
Deryagin \cite{dery} and has found broader application as
the Proximity Force Theorem \cite{pft}.  For this geometry, Eq. (1)
becomes
\begin{equation}
F(d)=2\pi R E(d)=2\pi {\pi^2\over 240}{1\over 3}{\hbar c\over d^3}
\end{equation}
where $R$ is the radius of curvature and $E(d)$ is the energy per
unit area that leads to the force in Eq. (1).  It should be noted
that the plate area does not enter into Eq. (2).  The two
recent experiments each used one convex and one flat plate.

\subsection{Torsion Pendulum Experiment}

Because I was directly involved in this experiment, I will give
an anecdotal account of it; specific details, schematics,
and data
can be found in \cite{lam}.  This experiment was started in
1994 as an undergraduate project. Dev Sen, a junior at the 
University of Washington at that time, was working with
our atomic physics group; he heard of the Casimir force in
his Electromagnetism course and asked if we might be able to
measure it.  With no knowledge of previous experimental work,
with the exception of the Sparnaay paper of 1958 \cite{spar2}
(which is routinely quoted
as ``the'' verification of Eq. (1)), I proposed an electrostatically
balanced hanging (Cavendish)
torsion pendulum technique; this is different from the horizontal
torsion balance used by van Silfhout \cite{van} (which, at the time,
we did not know about) particularly because the 
angular force constant $\alpha$ is at least two orders of magnitude smaller
for the hanging pendulum.  For short times (a few seconds of
averaging) our principal noise source was thermal fluctuations
(e.g., Brownian motion);
in this limit, the signal to noise scales as $1/\sqrt{\alpha}$.  
It appears that this noise was a major limitation of our experiment,
particularly when measuring the force at large plate separation. 

We constructed the
apparatus with junk found around the lab, and the total material
construction cost of the apparatus was around \$300.
The torsion pendulum fiber was about 60 cm in length; mounted on one
arm of the pendulum was a flat plate coated with a Cu/Au film, 
while the other arm served as one electrode
of a differential capacitor system that was used in
a feedback system to keep the torsion pendulum at a fixed angle.
The second plate used for the Casimir force measurement was mounted
on a precision mechanical mount and could be moved about 10 $\mu$m
by use of piezoelectric transducers.  
 
In our first attempts to measure the Casimir force, two flat
plates were used.  We were never able to properly align the
plates, and we eventually gave up when Dev graduated.  
Some months before I moved to Los Alamos, I decided to give
the apparatus one last try.  I had the idea to replace
one flat plate with a convex plate; I also improved the
vacuum and electronic circuitry.

In June of 1996 the apparatus became operational; when the two
plates were moved into close proximity, the increase in force 
over the residual electrostatic interaction
was easily evident.
Unfortunately, the apparatus was not stable enough to make 
a measurement; about 20 minutes after adjusting the distance
between the plates so that the separation at closest approach
would be of order 1 $\mu$m or less, the apparatus would drift
until the closest approach would increase to over 5 $\mu$m.
I eventually discovered the source of the drift:  the concrete
floor of the Physics Department basement would distort while I was
standing near the apparatus during adjustment, and the floor would
relax to its unstressed position over a 20 minute period.  
This effect was dealt with by standing in a different position,
rather far from the apparatus, while making the initial adjustments;
the experiment thereafter became physically (in an anatomical sense)
painful.

Data was taken over a one month period, after which the apparatus
was calibrated.  The final measurement was the radius of curvature
of the Cu/Au coated 
convex plate (lens) in which I made an error.  To make this measurement,
I used the angular deviation of a laser beam reflected from the
lens surface as the lens was translated normally to the beam.
Measurements near the center yielded $R=13.5\pm 0.5$ cm, while translations
across the entire lens yielded $R=11.3 \pm 0.1$ cm; it did not
occur to me that the lens might be aspheric (it was a high
quality, precision optical lens of unknown origin or application
that was coated with a thin layer of Au on top of a relatively thick Cu layer).
I assumed the latter
number was more accurate, and the data was described by Eq. (2)
with no corrections for finite conductivity; I mistakenly convinced myself
that the corrections should be less than 5\%.  At this point,
my work was rudely interrupted by my relocating to Los Alamos,
and I published my results in Physical Review Letters.

Subsequently, a number of theorists expressed surprise that no
correction due to finite conductivity was required \cite{eber}, and several
expressed an interest in doing a proper calculation.  None did
the calculation, so I figured out how to do it myself\cite{lam2}; I knew
that the plasma model correction to first order was not applicable
(see Sec. IIIa); being upwards of a 30\% correction,
none of the uncertainties in calibration were large enough to
encompass it.  However, the variation in the
measurement of $R$ was in the back of my mind, and in October
of 1997 I made a trip to Seattle and retrieved the lens.
Using a mechanical gauge, I measured $R=12.5$ cm in the region where
the Casimir force 
measurement was made.  The accurate finite conductivity correction
is small enough that its effect is indistinguishable from a calibration
error, or a mismeasurement of $R$.  In the end, my data did not agree with
the assumption that the film was pure Au, but was better described
by a pure Cu film \cite{lame}.  This result is not so surprising considering
that the properties of the film depend on preparation technique,
purity, and the possibility that the Au diffused into the Cu layer
significantly.  A more accurate calculation of the finite conductivity
effect would require a direct measurement of the complex permittivity
of the films.

In hindsight, it is remarkable that the torsion balance experiment,
which was intended as a demonstration,
worked as well as it did.  The improvement over previous measurements
is due to a number of factors, including the high sensitivity of
the hanging torsion pendulum and its lack of mechanical hysteresis, 
larger measurement distances so vibration and mechanical instabilities
were less important, improved piezoelectric transducers, and automated
data collection so that large amounts of data could be analyzed and
averaged.  

Much improvement over the present accuracy
obtained by this technique is unlikely.  The apparatus was rather unwieldy
with its enormous vacuum can and its susceptibility to tilt.  The length
of the torsion fiber might be significantly shortened, reducing
both the intrinsic sensitivity (bad) and sensitivity to
external perturbations (good); however
we must bear in mind that a factor of ten improvement in sensitivity
only extends the measurement distance by factor of about two.

\subsection{Atomic Force Microscopy Experiment}

In his 1989 review \cite{spar}, Sparnaay discusses the possibility of 
using atomic force microscopy (AFM) to measure the Casimir force; 
AFM had just been invented at that time \cite{binnig}.  It was
not until late 1998 that results from an AFM Casimir experiment
were reported by Mohideen and Roy \cite{moh}.

In this experiment, an Au/Pd $+$ Al coated, 0.3 mm polystyrene
sphere is attached to an AFM cantilever.  A similarly coated
optically polished sapphire plate was attached to a piezoelectric
transducer and brought near the sphere.  The attractive force
was determined by reflecting a laser beam from the cantilever
tip; the displacement of the laser beam on a pair of photodiodes
produced a difference signal proportional to the cantilever bending
angle.  

The sensitivity of the apparatus was such that the absolute force
could be determined with a fractional error of 1\% at $d=100$ nm,
and about 100\% at 900 nm.  

The use of AFM to measure the Casimir force might be a real
breakthrough; this is because the AFM technique is very stable
and reproducible.  Unfortunately, it is limited to
measurements short distance where there are significant theoretical
uncertainties in the interpretation of the data.  For the
first time testing Eq. (1) or its modification for real materials
to better than 1\% accuracy appears possible.  
As described in \cite{moh}, it is anticipated that a factor of
1,000 improvement in sensitivity appears possible, which would extend the
separation where the Casimir force can be measured to 1\%
accuracy to about 1 $\mu$m.  At this distance, the theoretical
uncertainties associated with the corrections for real materials,
as described in the next Section,
become much less important.

\section{Corrections}

\subsection{Imperfect Conductivity}

Equation (1) must break down when the plate separation is so
small that the mode frequencies being affected when $d$ is varied
are above the material resonance or plasma frequencies.  In the
case of a simple metal, the real part of the dielectric constant
can be approximated by
\begin{equation}\label{plas}
\epsilon'(\omega)=1-\omega_p^2/\omega^2
\end{equation}
where $\omega_p$ is the the plasma frequency and is proportional
to the effective free electron density in the metal.  It is
convenient to introduce the plasma wavelength, $\lambda_p=2\pi c/\omega_p$.
Corrections to Eq. (1), expanded in terms of $\lambda_p/d$, have
been calculated to first order by Haergraves \cite{har} and by
Schwinger et al. \cite{schw}, and to second order by Bezerra et al.\cite{bez}
For flat plates, the corrected force can be written in terms of a Eq. (1)
with a multiplicative factor,
\begin{equation}\label{pcor}
F'(d)=F(d)\left[1-{8\over 3\pi}{\lambda_p\over d}+{120\over 4\pi^2}\left(
{\lambda_p\over d}\right)^2\right].
\end{equation}
This equation is only valid for $\lambda_p/d<<1$; unfortunately, the Casimir 
force
is large enough to be accurately measured experimentally only in the
range  $\lambda_p/d\approx 1$ or larger.
We are also faced with the problem that Eq. (\ref{plas}) is only
approximate.

It is, however, possible to very accurately determine the attractive force
as a function of plate separation by numerical calculation, provided
we know its complex permittivity as a function of frequency:
\begin{equation}
\epsilon(\omega)=\epsilon'(\omega)+i\epsilon''(\omega)
\end{equation}
where $\epsilon',\ \epsilon''$ are real.  With this information,
the permittivity along the imaginary axis can then be determined by
use of the Kramers-Kronig relation,
\begin{equation}
\epsilon(i\xi)={2\over \pi}\int_0^\infty {x\epsilon''(x)\over x^2+\xi^2}dx +1.
\end{equation}
This can be used in the Lifshitz expression for the attractive 
force \cite{lif},
\begin{eqnarray}\label{lif}
F'(d)&=&{\hbar\over 2\pi^2c^3}\int_0^\infty\int_1^\infty p^2\xi^3\nonumber \\
&\ &\Bigg(\left[{[s+p]^2\over[s-p]^2}e^{2p\xi d/c}-1\right]^{-1}\nonumber\\
&+&\left[{[s+\epsilon(i\xi)p]^2\over 
[s-\epsilon(i\xi p)]^2}e^{2p\xi d/c} -1\right]^{-1}\Bigg)
dp\ d\xi
\end{eqnarray}
where $s=\sqrt{\epsilon(ix/p)-1+p^2}$.
The numerical calculation for the attractive force between Au, Al, and
Cu plates has been recently published \cite{lam2}, and significant 
deviations from Eq. (\ref{pcor}) were found.  In particular, for
Al with $d\approx 100$ nm\, Eqs. (\ref{pcor}) and (\ref{lif})
differ by about 5\%; one should note that including the third order
correction to Eq. (\ref{pcor}) worsens the deviation.  However,
these calculations should be considered in light of the notorious
variation of bulk and surface properties of materials due to
preparation technique, purity, etc. \cite{over,white}.

\subsection{Surface Roughness}

From the earliest experiments, it was realized that surface roughness
would lead to an increase of the apparent Casimir force and therefore
cause systematic errors in measurements aimed at verifying Eq. (1).
Such effects were observed by van Blokland and Overbeek \cite{over};
roughness has been discussed theoretically by van Bree et al. \cite{bree},
and more recently in \cite{mos}.

For high-quality optically polished surfaces, the roughness RMS amplitude $A$ 
is usually of order $A= 30$ nm or less.  For a $1/d^4$ attractive force,
the correction to Eq. (1) can be written
\begin{equation}
F'(d)\approx F(d)\left[1+4\left(A\over d\right)^2\right].
\end{equation}
The correction for the recent torsion balance experiment, at the point of
closest approach, is about 1\%, while for the AFM experiment,
it is about 30\%.

The roughness correction was derived in the context of a $1/d^4$
force law (this can be easily modified for the spherical plate
$1/d^3$ case).  However, the finite conductivity correction,
particularly as given by Eq. (\ref{pcor}), effectively has
terms containing $1/d^5$ and $1/d^6$.  In principle, the roughness
correction should be done for each power law separately, or the average
force determined from the accurate calculation, Eq. (\ref{lif}).
One should also bear in mind that the simple geometrical averaging
procedure isn't exactly correct; a complete treatment
would involved solving the appropriate electromagnetic rough boundary
problem.  However, the geometrical averaging is correct so long as the
period of the roughness is larger than the separation between the plates.

\subsection{Effect of Thin Films on the Plate Surfaces}

Either intentionally (Au evaporated onto an Al or Cu coated substrate)
or accidentally (formation of oxide layers) every Casimir force 
measurement has made use of mono- or multilayer coated plates.
The calculation of the force for a general film configuration has
been given by Spruch and Zhou \cite{spruch}.
The problem that I will consider here
is very simple so I will outline its solution in some detail.

A simple  geometry that illustrates the
effect of a thin material film is shown in Fig. 1; one of two identical 
perfectly conducting
flat plates is coated with a thin layer (thickness $a$) of a real
substance (Au, for example), and
the separation between the perfectly conducting surfaces is $d+a$.
This simplified problem will allow us to determine the qualitative effect
of a thin film.

Milonni has presented a complete calculation for the case of
materials with no absorption (\cite{mil}), 
and this can be easily adapted to
the multilayer case (this calculational technique
was first described in \cite{vank} and elaborated
upon in \cite{barash}).  Following Milonni, we consider the case
where the electromagnetic wave propagation vector in the
three materials (vacuum, film, perfect conductor)
\begin{equation}\label{vec}
K_i^2=k^2-\epsilon(\omega){\omega^2\over c^2}
\end{equation}
where $k$ is a real number, 
and $i=0,1,2$ with 0 ($\epsilon_0(\omega)=1$) representing the space
between the plates and 2 ($\epsilon_2(\omega)=\infty$)
the perfect conductor, and we require
$Re(K_i)\geq 0$.  For the case considered here, we must allow
for a complex $\epsilon_1$ in which case the $K_i$s can be complex.
In the Appendix, the use of the techniques
described by Milonni for the case of absorption
(complex $\epsilon$) is justified.

There are two type of solutions to the wave equation, one with
electric vector parallel to the surfaces (with arbitrary orientation
which we choose as the $y$ axis), $e_y(z)$ and one with electric vector 
perpendicular to the surfaces (along the $z$ axis), $e_z(z)$.  The wave equation
is 
\begin{equation}
{d\over dz} e_{y,z}(z)-K_i^2e_{y,z}(z)=0
\end{equation}
and the boundary condition for $e_z$ are (1) $de_z/dz$ and $\epsilon e_z$
are continuous, while for $e_y$ they are (2) $de_y/dz$ and $e_y$ are
continuous (at the conducting surfaces, $e_y=0$ and $de_z/dz=0$).  Ignoring
unphysical exponentially growing solutions, we have
\begin{eqnarray} 
e_{y,z}(z)&=& A(e^{K_0z}\mp e^{-K_0z})\ \ \ \ 0\leq z\leq d-a \nonumber\\
&=& Be^{K_1z}+Ce^{-K_1z}\ \ \ \ \ d-a\leq z\leq d\nonumber\\
de_z/dz&=&0\ \ \ \ \ \ z=d\nonumber\\
e_y&=&0\ \ \ \ \ \ z=d
\end{eqnarray}
where the $\mp$ sets $e_y=0$ or $de_z/dz=0$ at the
conducting boundary located at $z=0$.  
We therefore have two sets of linear equations involving
$A,B,C$ for the two cases.  
The condition for non-trivial solutions of these equations
is that the determinant of the coefficient matrix is zero, yielding
the following two expressions:
\begin{eqnarray}\label{1}
f_{y}&(&\omega,k,d)=0\nonumber\\
&=&{[(e^{2K_1a}+1)K_1+(e^{2K_1a}-1)K_0]\over
[(e^{2K_1a}+1)K_1-(e^{2K_1a}-1)K_0]}e^{2K_0d}-1
\end{eqnarray}
\begin{eqnarray}\label{2}
f_{z}&(&\omega,k,d)=0\nonumber\\
&=&{[(e^{2K_1a}-1)K_1+\epsilon_1(\omega)(e^{2K_1a}+1)K_0]\over
[(1-e^{2K_1a})K_1+\epsilon_1(\omega)(e^{2K_1a}+1)K_0]}e^{2K_0d}-1.
\end{eqnarray}
The zeroes $\omega_{ny,nz}(k,d)$ of 
$f_{y,z}$ determine the allowed mode ``eigen'' frequencies,

The zero point energy associated with the plates is determined by
assigning energy $\hbar\omega/2$ to each mode;
\begin{equation}\label{zsum}
E(d)=
\sum_{n,\vec k}\left[ \hbar\omega_{ny}(k,d)/2+\hbar\omega_{nz}(k,d)/2\right]
\end{equation}
(in general, the eigen-frequencies are complex, but the imaginary
parts cancel as discussed in the Appendix).
where the sum over $k$, in the continuum limit, becomes an integral,
\begin{equation}
\sum_{\vec k} \rightarrow \left({L\over 2\pi}\right)^2\int 2\pi kdk
\end{equation}
where $L$ is the transverse dimensions of the plate in the $x,y$ directions.
The theory of complex function can be used to evaluate the sum over
eigen-frequencies; specifically, according to the argument theorem 
\cite{boas,church},
\begin{equation}
{1\over 2\pi i}\oint_C{f'(z)\over f(z)}dz=N-P
\end{equation}
where $C$ is a closed path in the complex plane, $N$ and $P$ are the number of
zeros and poles within $C$, respectively, and the path is counterclockwise.
The argument theorem can be modified to give the sum of the zeroes and
poles:
\begin{equation}\label{csum}
{1\over 2\pi i}\oint_C z{f'(z)\over f(z)}dz=\left[\sum z_i\right]_{f(z_i)=0}-
\left[\sum z_i\right]_{f(z_i)=\infty}.
\end{equation}
Furthermore, $f'(z)/f(z)=d/dz(\log f(z))$.  The eigen-frequencies of
physical interest lie in the right half plane; integrating along
the imaginary axis from $\infty$ to $-\infty$
and closing the path with a semi-circle at infinity around the right
half plane (see \cite{mil} for details), and integrating by parts gives
\begin{equation}\label{energy}
E(d)={\hbar L^2 \over 8\pi ^2}\int_0^\infty k\ dk\int_{-\infty}^\infty
d\xi \left[\log g_y(\xi,k,d)+\log g_z(\xi,k,d)\right]
\end{equation}
where we have set $\omega=i\xi$, with $\xi$ real, 
$g_{y,z}(\xi,k,d)=f_{x,y}(i\xi,k,d)$, and
\begin{equation}
K_i=k^2+\epsilon_i(i\xi)\xi^2/c^2.
\end{equation}
Finally, we note that the poles of Eqs. (\ref{1}) and (\ref{2}) do not depend
on $d$ because it only enters in the multiplicative exponential;
therefore, Eq. (\ref{energy}) gives the zero point energy up to
an additive constant, while the force per unit area is given by
\begin{equation}
-{\partial\over \partial d} E(d)=-{\hbar\over 4\pi^2}\int_0^\infty k\ dk
\int_0^\infty d\xi K_0\left[{1\over g_y(\xi,k,d)}+{1\over g_z(\xi,k,d)}\right]
\end{equation}
where (possibly) non-physical $d$-independent terms are omitted.

$F(d)$ for a Au film 35 nm thick is shown in Fig. 2; $\epsilon_1(i\xi)$
was determined from tabulated optical constants as described in
\cite{lam2}.  

\section{Conclusion}

In the last few years, two new experiments to verify the existence
of the Casimir force between solid bodies have been performed and
have shown a new level of accuracy; in particular, atomic force
microscopy
technique offers great promise for testing Eq. (1) and its modifications
for the case of real materials to better than 1\% precision. 
In particular, if the sensitivity of the AFM experiment can
be increased substantially, the measurement region can be
extended to larger separations where the theoretical uncertainties
discussed above are substantially reduced.
Perhaps even the effect of finite temperature \cite{brown}
will be measurable in the not-too-distant future.

\section{Appendix}

As discussed by Milonni \cite{mil}, the contour integration technique used to
sum the eigen-frequencies is technically correct only if the
eigen-frequencies lie on the positive real axis.  There has been
much commentary on the extension of this technique to absorptive
materials (i.e., complex permittivity) in which
case the eigen-frequencies are complex.  Barash and Giszburg 
\cite{barash} introduce the idea of an auxiliary system to account for the
complex permittivity, with the fundamental eigen-frequencies real.
However, the contour integration method gives the correct answer
quite simply; the mathematics does not know about the auxiliary
system.  I would like to offer a  non-rigorous explanation
for why the technique works in the case of absorption.

First, for a generalized permittivity \cite{ll}, 
$\epsilon(-\omega^*)=\epsilon^*(\omega)$.  
Therefore, the eigen-frequencies for the general boundary
problem case occur in
pairs, $\omega=\pm\omega'+i\omega''$.  We further take the case
where $Re(K_i)$ are either all positive or all negative which
follows from continuation; as $\omega\rightarrow\infty$, all 
the $K_i$ become equal because $\epsilon_i(\omega)\rightarrow 1$.
In the case that $Re(K_i)>0$, representing exponentially damped
surface wave, the eigen-frequencies lie in the lower half plane;
therefore, $e^{-i\omega t}$ is damped exponentially in time.
For $Re(K_i)<0$, the eigen-frequencies lie in the upper half plane
and represent solutions growing exponentially in time and space.
Clearly, the contour integration method as described above should
not work without justification regarding branch cuts etc.

The way to see around this problem is instead of considering the
eigen-frequencies, one can consider the corresponding eigen-wavenumbers $K_i$,
and $K_0$ in particular.  The eigen-wavenumbers occur in complex
conjugate pairs
$K_0=K_0'\pm i K_0''$ (as can be seen
from Eq. (\ref{vec}) and the properties of $\epsilon(\omega)$ in
the complex plane) and  by definition the $K_i$ 
are in the right half plane for the exponentially damped solutions.
Furthermore, if we write the determinant function in terms of $K_0$,
and using the fact that $K_0 dK_0=-\omega d\omega/c^2$, we find
\begin{equation}\label{wcont}
-c^2\oint_C K_0 {f'(\omega(K_0))\over f(\omega(K_0)} d\ K_0
=\oint_C \omega(K_0)  {f'(\omega(K_0))
\over f(\omega(K_0))} d\ K_0
\end{equation}
and we can replace $d K_0$ by $d\omega$ because the path is arbitrary
in the complex plane (note that in the case of no absorption, the
eigenvalues for $K_0$ lie on the imaginary axis while those for
$\omega$ lie on the real axis, and the contour must be adjusted
accordingly).
Because the eigenvalues for $K_0$ occur in conjugate pairs, the left side of
Eq. (\ref{wcont}) is
real; therefore, the sum in Eq. (\ref{zsum}) is real, as can
be seen from Eq. (\ref{csum}) when the integral is taken along
the imaginary $\omega$ axis.

\vfill\eject
\begin{figure}
\caption{Simple geometry for determining the effect of a thin
film on a perfectly conducting plate}
\end{figure}

\bigskip
\begin{figure}
\caption{The numercially calculated effect of a 35 nm thick Au film
on a perfectly conducting surface.  Lower curve, no coating;
middle curve, Au film; upper curve, perfectly conducting film.
The Au film effect is of order 50\% of the perfectly conducting
film effect in the 100-200 nm range.}
\end{figure}

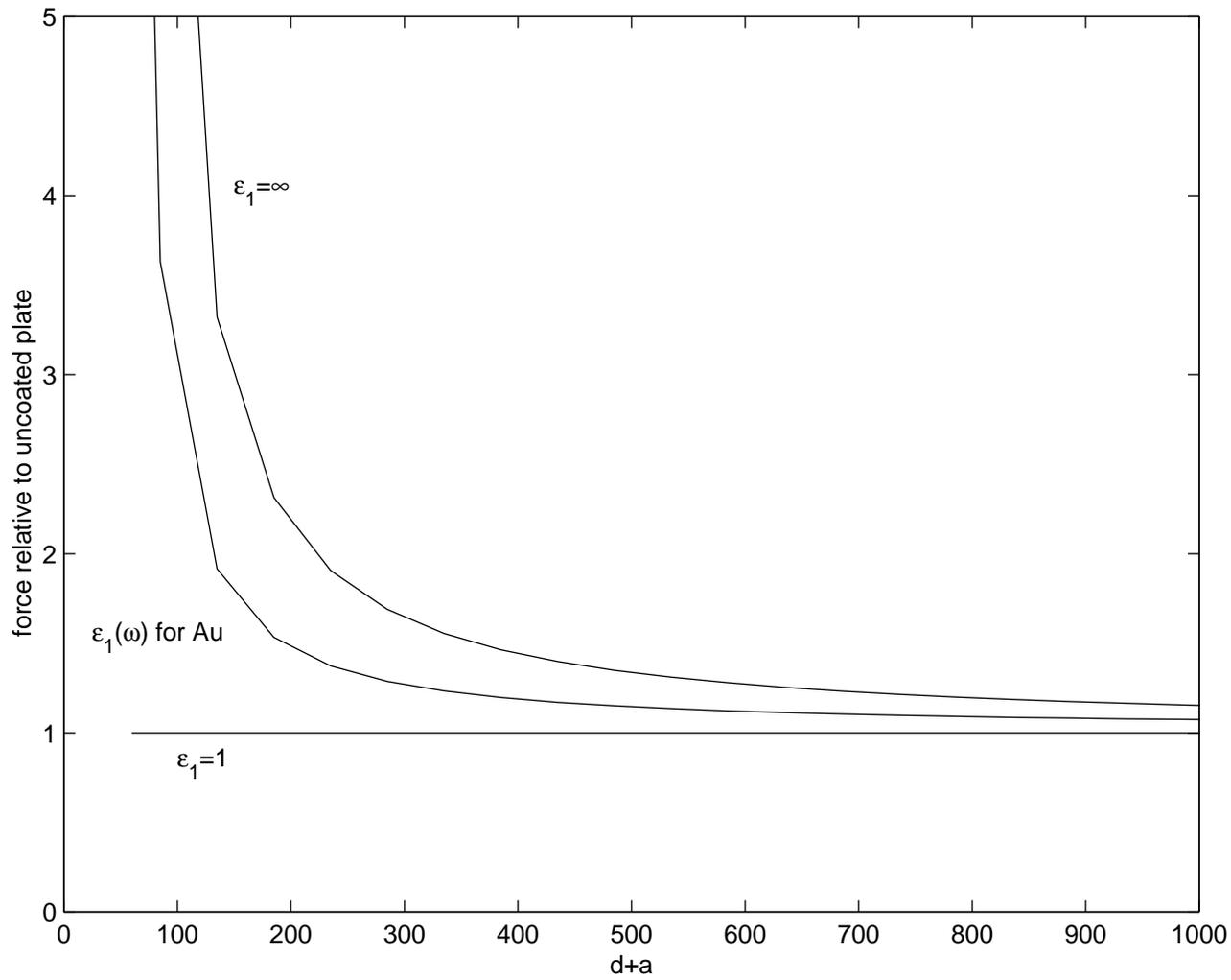
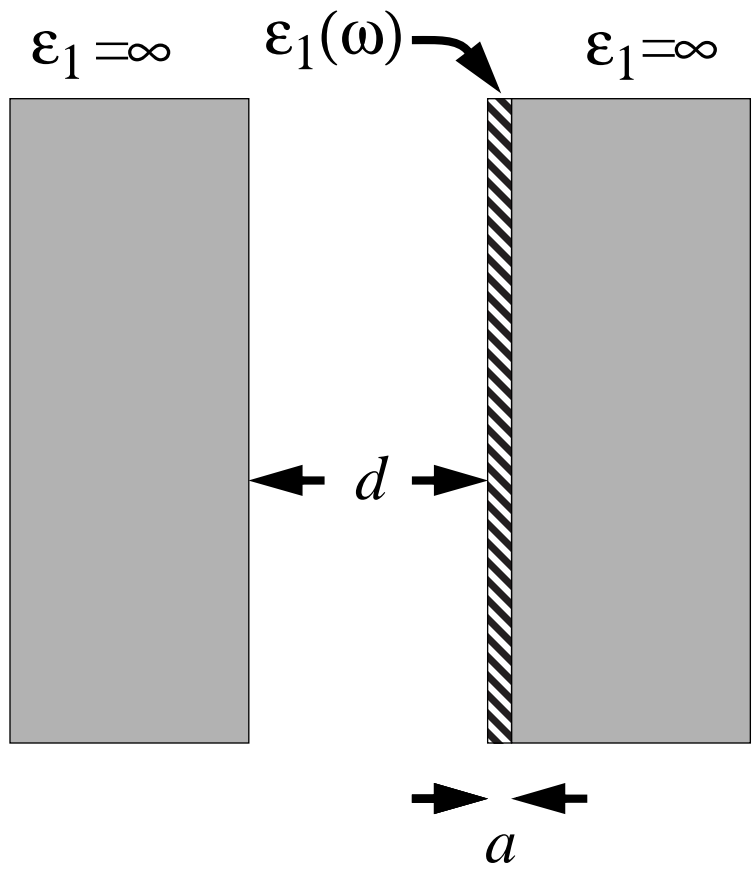
\end{document}